\documentclass[letterpaper]{article}
\usepackage{aaai20}
\usepackage{times}
\usepackage{helvet}
\usepackage{courier}
\usepackage[hyphens]{url}
\usepackage{graphicx}
\usepackage{graphicx}
\title{Advancing Artificial Intelligence and Machine Learning in the U.S. Government Through Improved Public Competitions}
\author{Ezekiel J. Maier\\
Booz Allen Hamilton\\
4747 Bethesda Ave.\\
Bethesda, MD 20814
}

\begin{document}

\maketitle
\begin{abstract}
In the last two years, the U.S. government has emphasized the importance of accelerating artificial intelligence (AI) and machine learning (ML)      within the government and across the nation. In particular, the National Artificial Intelligence Initiative Act of 2020, which became law on January 1, 2021, provides for a coordinated program across the entire federal government to accelerate AI research and application. The U.S. government can benefit from public artificial intelligence and machine learning challenges through the development of novel algorithms and participation in experiential training. Although the public, private, and non-profit sectors have a history of leveraging crowdsourcing initiatives to generate novel solutions to difficult problems and engage stakeholders, interest in public competitions has waned in recent years as a result of at least three major factors: (1) a lack of high-quality, high-impact data; (2) a narrow engagement focus on specialized groups; and (3) insufficient operationalization of challenge results. Herein we identify common issues and recommend approaches to increase the effectiveness of challenges. To address these barriers, enabling the use of public competitions for accelerating AI and ML practice, the U.S. government must leverage methods that protect sensitive data while enabling modelling, enable easier participation, empower deployment of validated models, and incentivize engagement from broad sections of the population.
\end{abstract}

\section{Introduction}
The White House, Congress, and Federal agencies recognize the benefits of artificial intelligence (AI) and machine learning (ML) and are accelerating AI and ML adoption. In addition to serving citizens more effectively, acceleration of AI and ML research and application is critical for the economic prosperity and national security of the U.S. (Schmidt et al. 2021). In particular, federal agencies are (1) modernizing infrastructure to support AI and ML development and operations; (2) adopting AI and ML solutions for improving and automating business processes; and (3) offering training to increase staff AI and ML awareness. For example, the U.S. Food and Drug Administration (FDA), through its Technology Modernization Action Plan (TMAP) and Data Modernization Action Plan (DMAP), is upgrading FDA’s technical infrastructure; building processes for innovative product development (such as AI and ML models); developing consistent, repeatable, and modern data management practices; and developing data science talent within the workforce via recruitment and retention activities, training, and knowledge sharing (FDA 2019) (FDA 2021). While these actions are advancing the use of AI and ML within the federal government, these steps alone are not sufficient to ensure that the most beneficial applications of AI and ML are prioritized and staff are AI-ready.

Federal agencies often offer training on informatics, data science, and data management via virtual self-paced on-demand courses, webinars, tutorials and documentation, and classroom-based learning (National Library of Medicine 2021). While this variety of approaches provides flexibility to learners, the lack of time constraints and instructor feedback may hinder course completion and reduce the ability of learners to convert gained knowledge to action. Augmenting current training options with experiential learning techniques will boost engagement and empower staff to seek and utilize AI and ML solutions. Experiential learning is employed in a variety of learning settings (e.g., medical and business school) to better engage students, enable collaboration and creativity, and achieve a better real-world understanding of a topic. AI-focused public challenges can provide similar experiential learning benefits, bridging the gap between self-paced virtual training and utilizing AI and ML.

The public, private, and nonprofit sectors have all utilized crowdsourcing to increase engagement, spur innovation, and solve real-world problems. In fact, the U.S. government has made a significant commitment to engaging citizens in voluntary crowdsourcing activities through its Challenge.gov platform which serves as a central source of government-wide challenges and prize competitions. Challenge.gov links to competitions on a variety of other federal government platforms including the National Aeronautics and Space Administration’s (NASA) Tournament Lab (Gustetic et al. 2015) and the FDA’s precisionFDA platform (Altman et al. 2016). Currently, Challenge.gov is hosting 29 active federal government challenges. In addition, 849 challenges were completed on Challenge.gov between 2010-2020. The Department of Health and Human Services (HHS),  NASA, and Department of Defense have been crowdsourcing leaders, each running more than 60 challenges since 2010. Moreover, more than 25 Challenge.gov hosted challenges have focused on AI and ML model development since 2018. In the private and non-profit sectors, notable crowdsourcing challenge platforms include Kaggle, InnoCentive, TopCoder, and the DREAM Challenges. Technically complex competitions, including AI, ML, and bioinformatics, are targeted toward industry, research communities, and educational institutions where they strengthen collaboration, engage new organizations and individuals, encourage innovation, supply opportunities for hands-on training, increase openness and availability of high-quality data sets and tools, and provide independent evaluations of tools and techniques.

Despite this commitment from public, private, and nonprofit sectors, and the significant gains in educational attainment and internet access, the interest and perceived effectiveness of crowdsourcing competitions is decreasing. Notably, the global gross enrollment ratio in tertiary education increased to 38\% in 2017 (UNESCO 2019), and there has been a 10\% annual increase in worldwide Internet users, topping out at more than 4.1 billion in 2019 (International Telecommunication Union 2019). Within the United States, 83.7 million adults, aged 25 and over, have achieved a bachelor’s degree or higher as of 2020 (United States Census Bureau 2021). In addition, Google Trends shows that worldwide web search interest in the term ”machine learning” has been at or near all-time highs since early 2019. Yet interest in “crowdsourcing” has decreased.  Google Trends shows a decrease by more than 50\% for the term “crowdsourcing” since peaking in late 2013. Moreover, Citizen Data Science is rated as entering the “Trough of Disillusionment” in the 2021 Gartner Hype Cycle for Machine Learning and Data Science. This practice paper reports on a novel initial exploratory analysis of U.S. government hosted public challenges, and describes opportunities to reinvigorate competitions by leveraging under utilized and unused approaches in the crowdsourcing community. 

\section{Barriers to the Long-Term Success of Crowdsourcing Competitions}

Since 2010, the U.S. government has invested significantly in crowdsourcing efforts. In addition to the innumerable person-hours spent organizing and running challenges, the government has allocated more than \$204 million dollars in prizes for the 878 completed and active challenges hosted on Challenge.gov. To increase the effectiveness of crowdsourcing competitions for advancing AI and ML literacy and applications, U.S. government organized challenges must better align with the expectations placed on them (Simula 2013). There are three major barriers that decrease the effectiveness of U.S. government AI and ML crowdsourcing challenges: (1) a lack of high-quality, high-impact data, (2) a narrow engagement focus on specialized groups, and (3) insufficient operationalization of challenge results.

\paragraph{Insufficient high-quality publicly available high-impact data.} In order to protect personally identifiable information, deidentified or synthetic data often is used in place of sensitive data. By using synthetic data instead of, for example, electronic health records or human genomes, the resulting models may be less applicable to real-world problems, and as such, may dissuade public engagement and discourage operationalization of developed AI and ML models.

\paragraph{Narrow engagement focus on specialized groups.} Ideally, crowdsourcing competitions would leverage the ”wisdom of the crowd”. However, public challenges often are organized for, and advertised to, relatively small, specialized groups, such as academic data scientists and bioinformaticians. While specialized knowledge is important, the diverse thinking that results from engagement of a wider audience can lead to new innovations and improved understanding of AI and ML through experiential learning.

\paragraph{Insufficient operationalization of challenge results.} The post-challenge phase is crucial for extracting knowledge from the challenge results and validating, improving, and operationalizing models. However, challenge sponsors, organizers, and participants often spend the majority of their focus on the modeling phase. Without additional focus on the post-challenge collaborative phase, challenges will not provide the benefits or have the impact that they are capable of.

\section{Approaches to Reinvigorate Crowdsourcing Competitions}

There are several approaches to evolve and increase the effectiveness of crowdsourcing challenges for advancing AI and ML use in the U.S. Government. These approaches include: model-to-data, autoML, design-a-thons, MLOps, and the introduction of novel incentives.

\paragraph{Model-to-data.} Popularized by the DREAM Challenges, model-to-data approaches can address data privacy concerns by evaluating models in a secure private computational environment that holds the underlying sensitive data. In this approach, participants develop and train their model on nonsensitive data, then containerize and submit their model for evaluation in the private computational environment (Ellrott et al. 2019). A distributed model-to-data framework is being used in the current COVID-19 EHR DREAM Challenge to enable development and evaluation of models that use electronic health records (EHRs) to predict patient specific risk for COVID-19 associated health outcomes. The U.S. government AI and ML challenge community should continue to adopt model-to-data approaches, enabling challenges that use high-quality, real, high-impact data and produce generalizable models.

\paragraph{AutoML.} Automated machine learning (AutoML) tools automate many of the steps in the machine learning pipeline, including feature engineering, model selection, model training, and model validation (Waring, Lindvall, and Umeton 2020). There are a number of vendors (e.g., Amazon Web Services (AWS), Google Cloud, Microsoft Azure, DataRobot) and open source tools (e.g., H2O, R, Python) that provide autoML tools. The U.S. government AI and ML challenge community should leverage autoML to expand access to challenges and increase efficiency. For example, beginner tracks of AI and ML challenges can be hosted on user-friendly point-and-click interfaces that simplify modeling-based decision making.

\paragraph{Design-a-thons.} Similar to hack-a-thons, design-a-thons engage a broad array of stakeholders to ideate possible solutions to real-world problems. Importantly, design-a-thons are welcoming to a broader audience by not requiring specialized subject matter or programming knowledge for participation. For example, the precisionFDA platform recently hosted the FDA New Era of Smarter Food Safety Low- or No-Cost Tech-Enabled Traceability Challenge to promote ideation and innovation of hardware, software, and advanced analytics solutions for enabling digital traceability along the entire food system. By requiring      PowerPoint and video presentations, rather than an implemented solution, this challenge enabled broader participation, leading to more than 90 submissions. The U.S. government AI and ML challenge community should leverage design-a-thons to engage employees and the public in the prioritization of AI and ML use cases.

\paragraph{MLOps.} Machine learning operations (MLOps) is a set of machine learning and DevOps practices for developing, deploying, and maintaining machine learning solutions, which includes model and data versioning, pipeline automation, testing, continuous integration and continuous delivery, and monitoring. The U.S. government AI and ML challenge community should adopt MLOps in the post-challenge phase to ensure that community developed models can be operationalized to benefit the government and public by being testable, transparent, scalable, secure, and reproducible. For example, the three top performing teams from the precisionFDA NCI-CPTAC Multi-omics Enabled Sample Mislabeling Correction Challenge participated in a collaborative post-challenge phase with the challenge organizers to (1) validate their computational methods for identifying and correcting sample mislabeling on independent datasets and (2) generate a single-best consensus pipeline. This consensus approach, named COrrection of Sample Mislabeling by Omics (COSMO), was developed and validated following MLOps considerations including scalability, reproducibility, and deployability via the use of Docker containerization and Nextflow (Yoo et al. 2021).

\paragraph{Novel incentives.} Novel incentives for top performance, such as fast-tracking pilots, partnerships, and contracts, will boost engagement while ensuring that clear steps are in place to reward winning models. For example, Artificial Intelligence Tech Sprints, organized by the National Artificial Intelligence Institute (NAII), award both monetary prizes and opportunities for partnership and piloting of selected prototypes (National Artificial Intelligence Institute 2020). Moreover, while only 3.3\% of the 432 challenges completed on the Kaggle platform from 2010-2020 utilized jobs as an incentive, there is an 80\% increase in the median number of participating teams as compared to the 70\% of Kaggle challenges that utilized monetary incentives.

\section{Conclusions and Discussion}
The U.S. government is leading a national initiative to accelerate AI/ML research and development, and upskill the workforce to enable AI/ML integrattion. Public crowdsourcing challenges have long been used as a tool for innovation and stakeholder engagement. For example, from 2006-2009 Netflix ran the Netflix Prize public competition, which offered a grand prize of \$1,000,000 to the top performing team that could improve the prediction of user ratings of films by 10\%. Through this public competition, Netflix was able to directly engage with over 40,000 registered teams that participated in the challenge, encourage advancements in the field of collaborative filtering (Koren and Bell 2015), and increase the public's awareness of Netflix and recommendation systems. Improved utilization of  public crowdsourcing competitions can provide the U.S. government with similar benefits, including AI/ML innovation and improved workforce AI-readiness. To achieve these benefits, U.S. government sponsored public challenges must democratize challenge participation (e.g., through autoML, design-a-thons, and novel incentives), enable validation on real-world data (e.g., via model-to-data), and focus on operationalizing high-performing validated models (e.g., using MLOps).

In addition to the crowdsourcing barriers and improvements discussed in this paper, more study is needed to identify and quantify the factors that influence the success of public competitions. More analysis of public challenges, such as those hosted by Challenge.gov and Kaggle, is needed to understand all predictive factors. To empower this analysis, the crowdsourcing community, including the Federal Community of Practice on Crowdsourcinig and Citizen Scienc (FedCCS), must improve the measurement and documentation of challenge outcomes. Measurement and documentation of outcomes, such as deployed models, scientific publications, and educational attainment will strengthen subsequent recommendations and ultimately increase the impact of federal AI/ML public challenges.

\section{References}

\setlength{\parskip}{0.5em}

\noindent Altman, R. B., Prabhu, S., Sidow, A., Zook, J. M., Goldfeder, R., Litwack, D., ... and Giacomini, K. M. 2016. A research roadmap for next-generation sequencing informatics. {\it Science translational medicine}  8(335): 335ps10-335ps10.

\noindent Ellrott, K., Buchanan, A., Creason, A., Mason, M., Schaffter, T., Hoff, B., ... and Saez-Rodriguez, J. 2019. Reproducible biomedical benchmarking in the cloud: lessons from crowd-sourced data challenges. {\it Genome biology} 20(1): 1-9.

\noindent Food and Drug Administration. 2019. FDA’s Technology Modernization Action Plan (TMAP). URL https://www.fda.gov/media/130883/download (accessed 10.31.21). 

\noindent Food and Drug Administration. 2021. Data Modernization Action Plan (DMAP). URL https://www.fda.gov/media/143627/download (accessed 10.31.21). 

\noindent Gustetic, J. L., Crusan, J., Rader, S., and Ortega, S. 2015. Outcome-driven open innovation at NASA. {\it Space Policy}, 34: 11-17.

\noindent International Telecommunication Union. 2019. Measuring digital development: Facts and figures 2019. {\it ITU Publications}. URL https://www.itu.int/en/ITU-D/Statistics/Documents/facts/FactsFigures2019.pdf (accessed 9.25.20). 

\noindent Koren, Y., Bell, R. 2015. Advances in collaborative filtering. {\it Recommender systems handbook}, 77-118.

\noindent National Artificial Intelligence Institute. 2020. Industry, academics invited to design AI tool to help Veterans. {\it VAntage Point}. URL https://blogs.va.gov/VAntage/77660/industry-academics-invited-design-ai-tool-help-veterans/ (accessed 9.25.20). 

\noindent National Library of Medicine. 2021. Training on Biomedical Informatics, Data Science, and Data Management. URL https://learn.nlm.nih.gov/documentation/training-packets/T000181112/ (accessed 10.31.21). 

\noindent Schmidt, E., Work, B., Catz, S., Chien, S., Darby, C., Ford, K., Griffiths, J.M., Horvitz, E., Jassy, A., Mark, W. and Matheny, J. 2021. National Security Commission on Artificial Intelligence Final Report. {\it National Security Commission on Artificial Intelligence}.

\noindent Simula, H. 2013. The rise and fall of crowdsourcing?. In {\it 2013 46th Hawaii International Conference on System Sciences}, 2783-2791. IEEE.

\noindent UNESCO. 2019. \#CommitToEducation. URL https://unesdoc.unesco.org/ark:/48223/pf0000370738 (accessed 9.25.20). 

\noindent United States Census Bureau. 2020. CPS Historical Time Series Visualizations. URL https://www.census.gov/library/visualizations/time-series/demo/cps-historical-time-series.html (accessed 8.13.21).

\noindent Waring, J., Lindvall, C., and Umeton, R. 2020. Automated machine learning: Review of the state-of-the-art and opportunities for healthcare. {\it Artificial Intelligence in Medicine} 101822.

\noindent Yoo, S., Shi, Z., Wen, B., Kho, S., Pan, R., Feng, H., ... and Zhang, B. 2021. A community effort to identify and correct mislabeled samples in proteogenomic studies. {\it Patterns} 2(5): 100245.

\bibliographystyle{aaai}
\end{document}